\documentstyle[aps,prl,array,epsf,floats]{revtex} 

\begin{document}
\twocolumn[{\hsize\textwidth\columnwidth\hsize\csname
@twocolumnfalse\endcsname
\draft
\title{
%%%%%%%%%%%%%%%%%%%%%%%%%%%%%%%%%%%%%%%%%%%%%%%%%%%%%%%%%%%%%%%%%%%%
%
        Synchronization of Short-Range Pulse-Coupled Oscillators 
%
%%%%%%%%%%%%%%%%%%%%%%%%%%%%%%%%%%%%%%%%%%%%%%%%%%%%%%%%%%%%%%%%%%%%
}
\author{ 
Jos\'e M.G. Vilar$^1$\cite{email}  
and 
\'{A}lvaro Corral$^2$\cite{email}
}
\address{ 
$^1$%
Departament de F\'{\i}sica Fonamental, 	
%Facultat de F\'{\i}sica, 
Universitat de Barcelona, Diagonal 647, E-08028 
Barcelona, Spain \\
$^2$%
The Niels Bohr Institute,        
University of Copenhagen,       
Blegdamsvej 17,                 
DK-2100 Copenhagen \O,                
Denmark
}
%\date{\today}
\date{July 29, 1998}
\maketitle  
\widetext
\begin{abstract}
\leftskip 54.8 pt
\rightskip 54.8 pt

We explore systems of pulse-coupled oscillators beyond the mean-field limit
[R.E.  Mirollo and S.H. Strogatz, 
{ SIAM J. Appl. Math. } {\bf 50}, 1645 (1990)]
by means of a manageable description which leads to a great
simplification of the dynamics.
Sufficient conditions for synchronization are exactly obtained 
for a ring with nearest-neighbor directed interactions, 
turning out to be the same as in the all-to-all case.
Our analysis shows new synchronization mechanisms occurring when
the interactions are local.

\end{abstract}

}]
\narrowtext

Nonequilibrium spatially extended dynamical systems, 
consisting of many degrees of freedom,
constitute one of the most frequent manifestations
in nature, which however still require
new approaches to be understood
\cite{Strogatz_95}.
Very remarkable in them
is the spontaneous emergence of order, i.e., self-organization,
by means of local, short-range interactions that cooperate to build  
correlations over all spatial and temporal scales \cite{Bak96}.
Among all forms of order in space and time,
synchronization and periodicity are the strongest ones, respectively,
and play a crucial role in many physical, chemical, and
biological systems 
\cite{Winfree,Kuramoto_book,Strogatz94}.

Systems behaving in this way are in fact 
populations of oscillators coupled between them;
in particular, outstanding examples 
make it clear that in most living systems the interaction between
units cannot be viewed as continuous in time but rather as
discontinuous and episodic,
in the form of sharp pulses from one oscillator to another.
Pacemaker cells in the heart \cite{Peskin}, 
neurons in the visual cortex \cite{Gray},
and fireflies that flash on and off
simultaneously in periodic intermittency
\cite{Buck}
provide striking illustrations of spontaneous 
synchronous evolution in nature, all of them with pulsatile 
interactions.

The discontinuous character of the coupling makes remarkably
difficult the mathematical analysis of these systems.
In the case of a network of all-to-all {\it pulse-coupled oscillators},
Mirollo and Strogatz, however,  were able to overcome the hardness
of the problem establishing a sufficient condition for
synchronization~\cite{Mirollo}.
Their result is based on the idea of absorption,
which means that once two oscillators get synchronized
they never desynchronize.
In this way, they proved that,  no matter the initial
conditions, synchronization 
arises cooperatively by a sequence of absorptions, 
in a self-organized process where no cell
leads the population but it is the whole that coordinates
the overall activity to achieve {\it self-synchronization}.

In contrast to the  Mirollo and Strogatz's scenario and
other situations with all-to-all, or mean-field, 
interactions~\cite{Kuramoto91,Abbott,vanVreeswijk,Gerstner},
for more realistic networks~\cite{Watts},
in which the firing of one unit does not
modify the state of all the others in the same way,
the concept of absorption is no longer valid and the
synchrony achieved by a subset of oscillators can be 
lost~\cite{Corral95_1,Bressloff}.
Thereby, in these cases
one would expect a weaker tendency to synchronize,
as occurs in other large networks of oscillators.
Numerical studies~\cite{Corral95_2},
however, indicate that the conditions for synchronization
in networks with reduced connectivity seem to be the same
as in the all-to-all case.
Despite this counterintuitive result was already conjectured
by Mirollo and Strogatz~\cite{Mirollo}, up to now there is no proof of 
synchronization  in any model
of pulse-coupled oscillators beyond the mean-field assumption.

In this Letter we address precisely the problem of establishing sufficient
conditions for synchronization to occur in networks of
pulse-coupled oscillators with local connectivity.
In this regard, we have exactly derived them for the
Mirollo and Strogatz's  model with nearest-neighbor
directed coupling in one dimension,
turning out to be the same as in the mean-field case \cite{Mirollo}. 
This kind of feed-forward connectivities arise frequently 
in the modeling of some layered structures of the brain
\cite{Hertz},
whereas in loop-like circuits,
they are relevant for cardiac arrhythmia
and the brain clock \cite{Glass}.

Consider a population of $N$ relaxation or integrate-and-fire oscillators,
where the units evolve in time integrating an excitatory driving
signal until the state variable of some of them reaches a threshold,
firing an instantaneous zero-width pulse  to their neighbors,
and then relaxing to a lower-value state, where the driving acts again.
In general, any system of oscillators
can be described in terms of the phases $\phi_i$ of its units
\cite{Gerstner,Diaz}.
For relaxation oscillators $\phi_i$ measures the normalized elapsed
time since the last firing of unit $i$ if there had been no interactions.
The time evolution of the units is then given by
%
%================================================================
\begin{eqnarray}
\nonumber
       \mbox{if } \phi_i < 1  \ \forall i  \Rightarrow
       \frac{d\phi_i}{dt}=\frac{1}{T},
       \ \ i=1,2, \dots N, \  \ \ \ \hspace{5em}
       \\
%       \mbox{ (integrate)},
%$$
%================================================================
%
\\
%whereas the coupling, % on a $1-d$ directed lattice,
%
%================================================================
%$$
\nonumber
       \mbox{if } \phi_i \ge 1  \Rightarrow
       \left\{
       \begin{array}{lll}
           \phi_i & \rightarrow & 0,\\
           \phi_{n(i)}  &\rightarrow & \phi_{n(i)}+\Delta(\phi_{n(i)})
           \ \ \mbox{ if } \phi_{n(i)} > 0,
       \end{array}
       \right.
%       \mbox{ (fire)}
%$$
\end{eqnarray}
%================================================================
%
where 
$T$ is the period of the oscillators and $\Delta(\phi)$
the so-called phase response curve, which gives the phase shift
of a unit when it gets a pulse from a neighbor.
Neighbors of unit $i$ are denoted by $n(i)$ and
for a one-dimensional directed lattice with coupling ``from left to right'', 
$n(i)=i+1$.
Periodic boundary conditions are taken assuming $N+1 \equiv 1$ and
$0 \equiv N$.

Notice that $T$ and $\Delta(\phi)$ are the same for all the units,
which therefore are identical and identically coupled.
Observe also that a unit at the reset point, $\phi=0$,
does not feel the incoming pulse;
this is a way to include the effects of absolute refractoriness.
In addition, as in  the Mirollo-Strogatz model, we only deal with
excitatory interactions, i.e., $\Delta(\phi)>0, \forall \phi$;
this makes the interaction process evolve in the form of 
avalanches.

%%%%%%%%%%%%%%%%%%%%%%%%%%%%%%%%%%%%%%%%%%%%%%%%%%%%%%%%%%%%%%%%%%
%\section{Proof}
%%%%%%%%%%%%%%%%%%%%%%%%%%%%%%%%%%%%%%%%%%%%%%%%%%%%%%%%%%%%%%%%%%

We will show that a sufficient condition for 
complete synchronization
is obtained when the phase response curve %$\Delta(\phi)$ 
is an increasing function of the phase,
i.e., $\Delta'(\phi)>0, \forall \phi $. 
Our proof of synchrony consists of two differentiated parts. 
The first one establishes that once
an oscillator has fired triggered by the firing of its neighbor,
both units will fire at the same time forever,
%two neighboring units fire at the same time,
%both units always remain firing at the same time 
even though they may lose their mutual synchrony between firings.
We call such a process a {\it capture}.
The second part shows how for almost all initial conditions
there is a capture for 
each pair of neighboring units.
%neighboring units fire  at least once at the same time.
In this way, 
as eventually every pair of neighbors are mutually captured,
the firing of one unit will ensure the firing of the rest
and the complete synchrony of the whole population will be achieved,
with all the units evolving with the same phase.

%%%%%%%%%%%%%%%%%%%%%%%%%%%%%%%%%%%%%%%%%%%%%%%%%%%%%%%%%%%%%%%%%%
%\subsection{First part}
%%%%%%%%%%%%%%%%%%%%%%%%%%%%%%%%%%%%%%%%%%%%%%%%%%%%%%%%%%%%%%%%%%

In order to prove our first claim,
let us consider two neighboring units,
labeled by $i$ and $i+1$,
which have fired at the same time.
Since both units have the same {\it free} time evolution,
they can only desynchronize when the $i$th one
gets a pulse;
when this happens, %the $i$th unit gets the pulse 
at time $a$, its phase $\phi^{a-}_{i}$ is instantaneously
increased by an amount $\Delta(\phi^{a-}_{i})$.
Here the superscript of the phase indicates time.
Notice that just before the pulse both phases are equal, i.e.,
\begin{equation}
     \phi^{a-}_{i}=\phi^{a-}_{i+1} \;.
\end{equation}
In contrast, just after the pulse 
the phase of the $(i+1)$th unit remains at the same value, in principle,
whereas the phase of the $i$th unit changes to
\begin{equation}
     \phi^{a+}_{i}=\phi^{a-}_{i}+\Delta(\phi^{a-}_{i}) \;.
\end{equation}
At this point, two different situations may occur depending on
whether the $i$th unit crosses the threshold as a consequence of
the received pulse or not.

If $\phi^{a-}_{i}+\Delta(\phi^{a-}_{i}) > 1$ the unit
crosses the threshold and fires, 
increasing then the phase of its right neighbor by an
amount $\Delta(\phi^{a-}_{i+1})$. 
Since  $\Delta(\phi^{a-}_{i+1})=\Delta(\phi^{a-}_{i})$,
the $(i+1)$th unit also crosses the threshold,
firing at the same time as the $i$th one.

If $\phi^{a-}_{i}+\Delta(\phi^{a-}_{i}) < 1$ the $i$th unit does not
reach the threshold. 
Therefore, after receiving the
pulse its phase
increases continuously in time until either it reaches the threshold
or it gets another pulse.
In the former case,
the unit fires at time $b=a+T[1-\phi^{a-}_{i}-\Delta(\phi^{a-}_{i})]$,
sending a pulse of strength $\Delta(\phi^{b-}_{i+1})$ to $i+1$. %its neighbor.
Notice that as
\begin{equation}
\phi^{b-}_{i+1}=1-\Delta(\phi^{a-}_{i})  \;,
\end{equation}
the phase of $i+1$ after the pulse will be
\begin{equation}
\phi^{b+}_{i+1} = 1-\Delta(\phi^{a-}_{i}) + \Delta(\phi^{b-}_{i+1}) \;.
\end{equation}
Assuming that
$\Delta(\phi)$ is an increasing function
of its argument, $\phi^{b+}_{i+1}$
is greater than $1$. Consequently the 
$(i+1)$th unit crosses the threshold, firing then
at the same time as its left neighbor.

Let us now discuss the case 
in which the $i$th unit receives two pulses.
Indeed, as the strength of the pulses is not fixed,
a unit can overtake its right neighbor;
this is what makes a unit get more than one pulse
while its phase is evolving between $0$ and $1$.
For the sake of generality we also consider that 
a set of connected units on the left of $i$ also
get two pulses, except the
leftmost unit of this set, denoted by $L$, which only gets a single pulse. 
The existence of a unit receiving only a single pulse
is guaranteed by the periodic conditions and by the fact
that at least $i+2$
gets only one pulse before $i+1$  can reach the
threshold.
After sending the first pulse to unit $j$, 
$j-1$ is delayed with respect to $j$
and consequently
\begin{equation}
\phi^{a_{j-1}}_{j-1} < \phi^{a_{j-1}}_{j} \;,
\;\; \forall j \mbox{~~s.t.~~} L < j \le i \;,
\end{equation}
where the superscript ${a_{j-1}}$ indicates any time between
the first firing of the $(j-1)$th unit and the time
in which $j-1$ gets its first pulse in this cycle.
In order for a unit $j$ to get the second pulse,
the second firing of its left neighbor must occur
before the second firing of $j$.
The sequence of firings
of this set is then from left to right, starting on the $L$th
unit.
Therefore, just before and after the $L$th unit gets
its only pulse at time $a$, its phase must be smaller and greater,
respectively, than the
corresponding  ones of the remaining units in the group, i.e.,
\begin{equation}
    \phi^{a-}_{L} <
    \phi^{a-}_{j} <
    \phi^{a-}_{L} + \Delta(\phi^{a-}_{L}) 
 \;, \;\; \forall j \mbox{~~s.t.~~}  L<j \le i+1 \;.
\end{equation}
Once the $L$th unit reaches the threshold
at time $b=a+T[1-\phi^{a-}_{L}-\Delta(\phi^{a-}_{L})]$, all units
from $L$ to $i+1$ fire because
\begin{eqnarray}
    \nonumber
    \phi^{b+}_{j} = 
      1
    +\phi^{a-}_{j}                             % change !!!!
    -\phi^{a-}_{L}                             % change !!!!
    +\Delta[\phi^{b-}_{j}] 
    -\Delta[\phi^{a-}_{L}]
    > 1 \;,
    \\
    \;\; \forall j \mbox{~~s.t.~~} L<j \le i+1 \;,
\end{eqnarray}
since
$\Delta'(\phi)>0$, $\forall \phi$
and
$\phi^{b-}_{j} > \phi^{a-}_{L} \;, \;\; \forall j \mbox{~~s.t.~~} L<j \le i+1$.

Our analysis then indicates that once
two neighboring units fire at the same time,
they always remain firing at the
same time if the left unit only gets one or two pulses.
The case in which the left unit gets three or more consecutive pulses
is not allowed in this process because a unit reaches the threshold and
fires when it gets the second pulse, as we have shown.

\begin{figure}[t]
\centerline{
\epsfxsize=7.0cm 
\epsffile{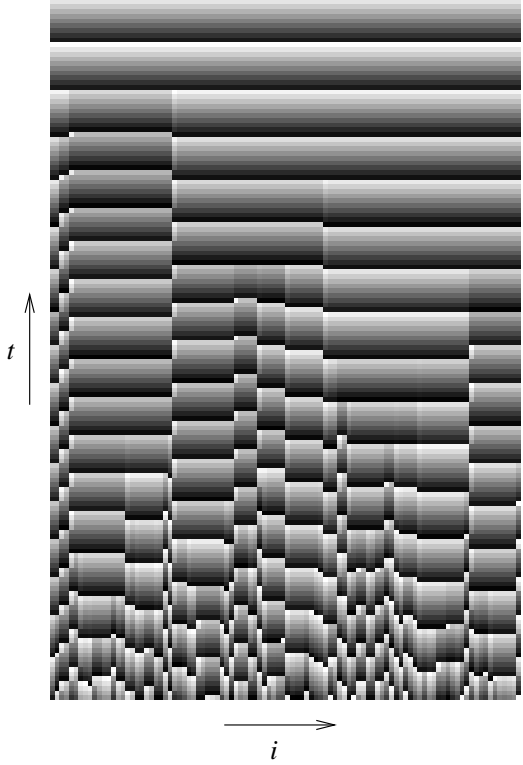}
}

\caption[a]{\label{fig1}
Spatiotemporal evolution of a ring of 100 oscillators for
$\Delta(\phi)=\varepsilon+\gamma\phi$, with $\varepsilon=0.01$
and $\gamma=0.3$.
Black and white colors stand  for minimum and
maximum values of the phase $\phi_i$, respectively.
Notice that the coupling is ``from left to right''.
}
\end{figure}

\begin{figure}[t]
\centerline{
\epsfxsize=7.5cm 
\epsffile{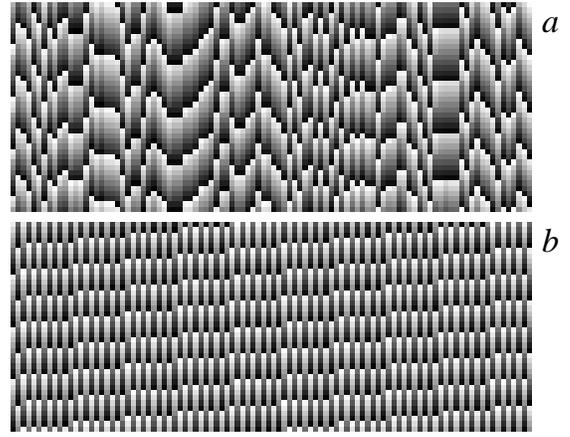}
}

\caption[a]{\label{fig2}
(a) Same situation as in Fig. \ref{fig1} but $\varepsilon=0.1$
and $\gamma=0$.
(b) Same situation as in Fig. \ref{fig1} but $\varepsilon=0.7$
and $\gamma=-0.1$.
}
\end{figure}

%%%%%%%%%%%%%%%%%%%%%%%%%%%%%%%%%%%%%%%%%%%%%%%%%%%%%%%%%%%%%%%%%%%
%\subsection{Second part}
%%%%%%%%%%%%%%%%%%%%%%%%%%%%%%%%%%%%%%%%%%%%%%%%%%%%%%%%%%%%%%%%%%%

Concerning our second statement, 
we now demonstrate that the set ${\cal A}^0$ of initial
conditions of an arbitrary oscillator
for which it will never fire at the same time
as its left neighbor has Lebesgue measure zero.
To be precise, for a unit $i+1$ 
and given an initial configuration of the rest of the population,
the set ${\cal A}^0_{i+1}$ is defined
as
\begin{eqnarray}
\nonumber
    {\cal A}^0_{i+1} \equiv
  { \bf  \{} \phi^0_{i+1} \mbox{ s.t. $i$ will not capture $i+1$ and}
\\
             \mbox{$i$ will not be captured 
                   in an avalanche triggered by $i+1$,}
\nonumber
\\ 
             \mbox{provided that the rest of the population}
\nonumber
\\
             \mbox{is in an initial state $\{\phi_j^0  \}_{j \ne i+1}$ }  
 { \bf \}}.
\nonumber
\end{eqnarray}
In particular, if $\phi^0_{i}=1$,
${\cal A}^0_{i+1}  \subset
            \{\phi^0_{i+1} $ s.t. 
                 $ 0 < \phi^0_{i+1} < 1-\Delta(\phi^{\star}) \}$,
where $\phi^{\star}$ is defined by $\phi^{\star}+\Delta(\phi^{\star})\equiv1$;
otherwise, $i$ captures $i+1$ immediately.
In essence, the proof shows that for almost all the initial 
conditions of a unit,
and irrespective of the state of the rest,
it cannot be confined forever 
in the interval $(0,1-\Delta(\phi^{\star}))$
receiving an infinite number of pulses from its neighbor.

Due to the characteristics of the dynamics,
the time evolution of any unit
can be decomposed into two parts.
On the one hand, the free time evolution in which the unit does
not get any pulse is given by
\begin{equation}
      \phi(t') \equiv F^{t'-t}[\phi(t)]=
      \left(\phi(t)+{t'-t\over T}\right)
      \; \mbox{mod} \; 1 \; ,
\end{equation}
where $t$ and $t'$ are times between two consecutive pulses.
Notice that time is now indicated by the argument of the phase.
On the other hand, the dynamics of the unit is coupled to
its neighbors. Thereby, when its left neighbor fires the unit
gets a pulse and increases its phase by an amount $\Delta(\phi)$, provided
that it is neither brought to the threshold 
nor refractory, which, by definition, is always fulfilled for the
time evolution of the set ${\cal A}^0$.
Then, 
the new phase just after the firing is
\begin{equation}
      \phi^{n+1}(t) \equiv h[\phi^n(t)]=\phi^n(t)+\Delta[\phi^n(t)] \;,
\end{equation}
with the superscript $n$ indicating now the number of pulses
received by the unit.

It can be seen that
the measure $\mu$ of the time evolution
of the set ${\cal A}^0$ remains invariant during the free time evolution 
by using that
\begin{equation}
       {d F^{t'-t}(\phi) \over d \phi}=1
    \ \ \mbox{for} \  0 \le \phi < 1 \;.
\end{equation}
Indeed, if ${\cal A}(t')$ denotes the state of the set ${\cal A}^0$
after a time $t'$,
\begin{eqnarray}
\nonumber
      \mu[{\cal A}(t')]  = 
      \int_{{\cal A}(t')} d\mu[\phi(t')] 
% & = & \int_{{\cal A}(t')} d\mu\{F^{t'-t}[\phi(t)]\} 
   =  \int_{{\cal A}(t)} {d F^{t'-t}[\phi(t)] \over d \phi(t)} d\mu[\phi(t)] \\
   =  \int_{{\cal A}(t)} d\mu[\phi(t)] = \mu[{\cal A}(t)] \;.
\nonumber
\end{eqnarray}
In contrast, when the unit gets a pulse,
the measure evolves
in such a way that the inequality 
\begin{equation}
\mu({\cal A}^{n+1})
\ge  \left(1+\tilde{\Delta'}\right)\mu({\cal A}^n)
\end{equation}
holds,
with $\tilde{\Delta'}\equiv\min[\Delta'(\phi)]$
and ${\cal A}^{n+1}$ being the resulting of the evolution
of the set ${\cal A}^0$ after the  $n+1$ pulses.
This follows directly from  both
\begin{equation}
\mu({\cal A}^{n+1}) =  \int_{{\cal A}^{n+1}}d\mu(\phi^{n+1}) 
 =  \int_{{\cal A}^n} { d h(\phi^n)\over d \phi^n } d\mu(\phi^n)
\end{equation}
and $  d h(\phi^n) / d \phi^n  =  1 + \Delta'(\phi^n) 
   \ge 1 + \tilde{\Delta'} $.
If $\Delta(\phi)$ is an increasing function of
its argument, its derivative is positive and then
$1+\tilde{\Delta'}$ is greater than 1.
In consequence, if the unit always gets the pulse
at $\phi \in (0,1-\Delta(\phi^{\star}))$, when $n$ goes to infinity
the inequality
\begin{equation}
\mu({\cal A}^0) \le { \mu({\cal A}^{n+1}) 
\over (1+\tilde{\Delta'})^{n+1}}
\end{equation}
implies that the set ${\cal A}^0$  has zero measure
since $\mu({\cal A}^{n+1})$ is bounded by $1-\Delta(\phi^{\star})$.

In other words, the flow in phase space remains incompressible
during the driving;
however, if $\Delta'(\phi) > 0$, $\forall \phi$,
the coupling acts in a more complicated way,
expanding the volume elements with each firing, except in the case
of a capture, where the volume contracts.
This is the process which eventually dominates,
leading the population to the complete-synchronization attractor.
In addition, it is easy to show that this condition is fully equivalent
to the one obtained by Mirollo and Strogatz \cite{Diaz}.

To illustrate the essentials of the
emergence of synchrony in systems with explicit spatial dependence,
in Fig. \ref{fig1} we have depicted the results obtained from 
numerical simulations for a directed ring when the condition
$\Delta'(\phi) > 0$, $\forall \phi$, is satisfied.
This figure shows the time evolution
towards the synchronized attractor,
starting from a random initial condition. It is evidenced how
the complete synchrony of the whole population is achieved
through a coarsening process in which the
domains of captured oscillators grow until
eventually all the units are evolving with the same phase.
In contrast, in Fig. \ref{fig2} we show the time evolution
for two situations wherein synchronization does not appear.
Fig. \ref{fig2}a corresponds to a case in which $\Delta'(\phi) = 0$,
$\forall \phi$.
In this situation, starting from a random initial condition,
perpetual disorganization occurs;
the system remains evolving periodically in time
in a spatially disordered state.
The situation shown in Fig. \ref{fig2}b corresponds to
a typical pattern reached after a long transient (not displayed)
for $\Delta'(\phi) < 0$,
$\forall \phi$. In this case, synchronization does not occur although
the final pattern looks more regular than that of Fig. \ref{fig2}a.

In summary, we have established, for the first time,
sufficient conditions for synchronization in a general model
of pulse-coupled oscillators with local interactions,
ensuring that, on a directed ring, they are the same
as in the mean field case.
The simple model of rhythmic oscillations in biological systems
we have presented illustrates the fundamentals of emergence
of organization in nonequilibrium complex systems led by
local rules, in contrast to previous work
devoted only to global coupling.
Our analysis, then, elucidates new mechanisms taking place
to attain synchrony when the spatial structure
is explicitly considered \cite{Stewart}.

We enjoyed very much the discussions with Steven Strogatz
during the XV Sitges Conference,
whose organization is also acknowleged by A.C. for providing
financial support to attend it.
J.M.G.V.'s work is financed by a grant of the CIRIT and
DGICYT's research program No. PB95-0881.

%%%%%%%%%%%%%%%%%%%%%%%%%%%%%%%%%%%%%%%%%%%%%%%%%%%%%%%%%%%%%%%%%%%
%%%%%%%%%%%%%%%%%%%%%%%%%%%%%%%%%%%%%%%%%%%%%%%%%%%%%%%%%%%%%%%%%%%


\begin{references}

%%%%%%%%%%%%%%%%%%%%%%%%%%%%%%%%%%%%%%%%%%%%%%%%%%%%%%%%%%%%%%%%%%%
%%%%%%%%%%%%%%%%%%%%%%%%%%%%%%%%%%%%%%%%%%%%%%%%%%%%%%%%%%%%%%%%%%%



\bibitem[*]{email}
E-mail addresses: {\tt vilar@ffn.ub.es}, {\tt corral@nbi.dk}.


\bibitem{Strogatz_95}
S.H. Strogatz,
{ Nature (London) } {\bf 378}, 444 (1995).

\bibitem{Bak96}
P. Bak,
{\it How Nature Works} (Copernicus, Springer-Verlag, New York, 1996).


\bibitem{Winfree} 
A.T.  Winfree, 
{\it The Geometry of Biological Time}
(Springer-Verlag, New York, 1980).

\bibitem{Kuramoto_book} 
Y. Kuramoto, 
{\it Chemical Oscillations, Waves, and Turbulence} 
(Springer-Verlag, Berlin, 1984).

\bibitem{Strogatz94}
S.H. Strogatz,
in {\it Frontiers in Mathematical Biology},
edited by S.A. Levin,
Lecture Notes in Biomathematics, Vol. 100, 122
(Springer-Verlag, Berlin, 1994);
%
%\bibitem{Wiesenfeld}
K. Wiesenfeld {\it et al.}, % P. Colet, and S.H. Strogatz,
Phys. Rev. Lett. {\bf 76}, 404 (1996).

\bibitem{Peskin} 
C.S.  Peskin, 
{\it Mathematical Aspects of Heart Physiology}, 
(Courant Institute of Mathematical Sciences, 
New York University, New York, 1975), p. 268.

\bibitem{Gray}
%C.M. Gray and W. Singer, 
%Proc. Natl. Acad. Sci. USA {\bf 86}, 1698 (1989);
C.M. Gray {\it et al.}, % P. K\"onig, A.K. Engel, and W. Singer, 
{ Nature (London) } {\bf 338}, 334 (1989).

\bibitem{Buck} 
J. Buck and E. Buck,
Sci. Am. {\bf 234}, 74 (1974).

%\bibitem{Smith} 
%H.M. Smith, 
%{ Science} {\bf 82}, 151 (1935).

\bibitem{Mirollo} 
R.E.  Mirollo and S.H. Strogatz, 
{ SIAM J. Appl. Math. } {\bf 50}, 1645 (1990).

\bibitem{Kuramoto91}
%\bibitem{5_p50d.15} 
         Y. Kuramoto, 
         { Physica D} {\bf 50}, 15 (1991);
%
%\bibitem{5_pre49.2668} 
         C.-C. Chen, 
         { Phys.  Rev. E} {\bf 49}, 2668 (1994);
% 
%\bibitem{5_prl74.4189} 
         S. Bottani, 
         { Phys. Rev. Lett. } {\bf 74}, 4189 (1995);
%         
%\bibitem{5_Bottani2}
%         S. Bottani, 
         { Phys.  Rev. E} {\bf 54}, 2334 (1996).
         
\bibitem{Abbott}
%\bibitem{5_pre48.1483} 
         L.F.  Abbott and C. van Vreeswijk, 
         { Phys.  Rev. E} {\bf 48}, 1483 (1993);
%
%\bibitem{5_n4.259} 
A.  Treves, { Network } {\bf 4}, 259 (1993);
%
%\bibitem{5_prl71.1280}
M. Tsodyks {\it et al.}, % I. Mitkov, and H. Sompolinsky, 
{ Phys. Rev. Lett.}
{\bf 71}, 1280 (1993).

\bibitem{vanVreeswijk}
%\bibitem{5_jcn1.313} 
C.  van Vreeswijk {\it et al.},  %L. F.  Abbott, and G. B.  Ermentrout,
{ J.  Comp.  Neur.} {\bf 1}, 313 (1994);
%
%\bibitem{5_prl74.1570}
U. Ernst {\it et al.},  % K. Pawelzik, and T. Geisel, 
{ Phys. Rev. Lett.}
{\bf 74}, 1570 (1995).

\bibitem{Gerstner} 
W.  Gerstner, { Phys. Rev. E} {\bf 51}, 738
(1995).

\bibitem{Watts}
%For an alternative point of view, see
D.J. Watts and S.H. Strogatz,
{ Nature (London) } {\bf 393}, 440 (1998).


\bibitem{Corral95_1}
%It is worth mentioning the close connection between
%short-range pulse-coupled oscillators and SOC models
%of earthquakes, see
%J.J. Hopfield,
%Phys. Today {\bf 47}, No. 2, 40 (1994);
A. Corral {\it et al.},
%, C.J. P\'erez, A. D\'{\i}az-Guilera, and A. Arenas,
%"Self-Organized Criticality and Synchronization in a Lattice Model
%of Integrate-and-Fire Oscillators."
{ Phys. Rev. Lett.} {\bf 74}, 118 (1995);
%C.J. P\'{e}rez, A. Corral, A. D\'{\i}az-Guilera, 
%K. Christensen, and A. Arenas,
%%``On Self-Organized Criticality and Synchronization in Lattice Models
%%of Coupled Dynamical Systems'',\\
%{ Int. J. Mod. Phys. B} {\bf 10}, 1111 (1996);
A. Corral {\it et al.},
%, C.J. P\'{e}rez, and A. D\'{\i}az-Guilera,
%"Self-Organized Criticality Induced by Diversity" 
{ Phys. Rev. Lett.} {\bf 78}, 1492 (1997).  

\bibitem{Bressloff}
P.C. Bressloff {\it et al.},
{ Phys. Rev. Lett.} {\bf 79}, 2791 (1997);  
P.C. Bressloff and S. Coombes,
{ Phys. Rev. Lett.} {\bf 80}, 4815 (1998).

\bibitem{Corral95_2} 
A. Corral {\it et al.},  
% C.J. P\'{e}rez, A. D\'{\i}az-Guilera, and A. Arenas,  
%"Synchronization in a Lattice Model of Pulse-Coupled Oscillators".
{ Phys. Rev. Lett.} {\bf 75}, 3697 (1995). 

\bibitem{Hertz}
J. Hertz, A. Krogh, and R.G. Palmer,
{\it Introduction to the Theory of Neural Computation}
(Addison-Wesley, Redwood, 1991).

\bibitem{Glass}
H. Ito and L. Glass,
Physica D {\bf 56}, 84 (1991);
%\bibitem{McCrone}
J. McCrone, New Scientist {No. 2106}, 52 (1997).

\bibitem{Diaz}
A. D\'{\i}az-Guilera {\it et al.},
%, A. Arenas, A. Corral, and  C.J. P\'{e}rez,
%``Stability of Spatio-Temporal Structures in a Lattice Model
%of Pulse-Coupled Oscillators'',\\
{ Physica D} {\bf 103}, 419 (1997). 

\bibitem{Stewart}
I. Stewart, 
{ Nature (London) } {\bf 350}, 557 (1991).



\end{references}
\end{document}